\documentclass[12pt]{article}
\begin{document}
\newcommand \be  {\begin{equation}}
\newcommand \bea {\begin{eqnarray} \nonumber }
\newcommand \ee  {\end{equation}}
\newcommand \eea {\end{eqnarray}}

\title{PRICE IMPACT}
\author{J. P. Bouchaud, Capital Fund Management}
\maketitle

\section{What is price impact?}

{\it Price impact} refers to the correlation between an incoming order (to buy or to sell) and the subsequent price change. That a buy trade should
push the price up seems at first sight obvious and is easily demonstrated empirically, as we will review below. It is also a dour reality for
traders for whom price impact is tantamount to a cost: their second buy trade is on average more expensive than the first because of their own impact (and
vice-versa for sells). Monitoring and controling impact is therefore one of the most active and rapidly expanding domains of research within trading firms. The most important questions are related to the volume dependence of impact (do larger trades impact prices more?), and the temporal behaviour of impact (is the 
impact of a trade immediate and permanent, or is there some lag dependence of the impact?).

A moment of reflection suggests that the interpretation of price impact is far from trivial, and may even lead to contradictions -- isn't a transaction a fair deal between a buyer and a seller? So why is there price impact? Three distinct possibilities come to mind:

\begin{enumerate}
\item
{\it Agents successfully forecast short term price movements and trade accordingly.} This can result in measurable correlation between trades
and price changes, even if the trades by themselves have absolutely no effect on prices at all. If an agent correctly forecasts price movements and
if the price is about to rise, the agent is likely to buy in anticipation of it. But in this framework `noise induced' trades, based on no information
at all, should have no price impact.
\item
{\it The impact of trades reveal some private information.}  The arrival of new private information causes trades, which cause other agents to update their valuations, leading to a price change. But if trades are anonymous and there is no easy way to distinguish
informed traders from non informed traders, then all trades will impact the price since other agents will believe that a fraction of these trades
might contain some private information. 
\item
{\it Impact is a statistical effect due to order flow fluctuations.}  Imagine for example a completely random order flow process, that leads to a certain order book dynamics (see, e.g. Farmer et al. 2005 for a description of such models). Conditional to an extra buy order, the price will on average move up if everything else is kept constant. Fluctuations in supply and demand can thus be completely random, unrelated to information, still a well defined notion of price impact will emerge. In this case impact is a completely mechanical -- or better, statistical -- phenomenon.  
\end{enumerate}

All three of these mechanisms result in some ``price impact'', i.e. a positive correlation of trading volume and price movement, but they are
conceptually very different. In the first two pictures, as emphasized in Hasbrouck (2007), ``orders do not {\it impact} prices. It is more accurate to say that orders {\it forecast} prices.'' In the second picture, price impact is of paramount importance to understand how information gets incorporated into prices. If some traders really have an information on the ``true'' price at some time in the future, then the observation of an excess of buy trades allows the market to guess that the price will move up and to change the quotes accordingly (see {\bf Glosten-Milgrom model, Kyle model} for explicit implementations). Thus while on one hand market impact is a friction, it should also be viewed as the mechanism that allows prices to adjust to new information.  

But if the third, mechanical, interpretation is correct, correlation between price changes and order flow is a tautology. If prices move only because of trades, ``information revelation'' may merely be a self-fulfilling prophecy which would occur even if the fraction of informed traders is zero. Possible differences between these pictures may come about in the volume dependence and temporal behavior of impact, which we will discuss below.

\section{Linear and permanent impact: The Kyle model}

The simplest assumption is that impact is both linear in the traded volume and permanent in time. These assumptions can be partly justified within the
Kyle model (Kyle, 1985), where an insider trader and noise traders submit orders that are cleared by a Market Maker (MM) every time step $\Delta t$. In this model, the price adjustment rule $\Delta p$ of the MM must be linear in the total signed volume $\epsilon v$, i.e:\footnote{$\epsilon = +1$ if the volume of
buys $v_b$ is larger than the volume of sells $v_s$ in the time interval $\Delta t$, and $\epsilon=-1$ in the opposite case; and $v = |v_b-v_s|$.}
\begin{equation}
\Delta p = \lambda \epsilon v,
\end{equation}
where $\lambda$ is a measure of impact, and is inversely proportional to the liquidity of the market. This price adjustement is furthermore {\it permanent}, i.e., the price change between time $t=0$ and $t=T=N \Delta t$ is:
\begin{equation}\label{evolution}
p_T = p_0 + \sum_{n=0}^{N-1} \Delta p_n = p_0 + \lambda \sum_{n=0}^{N-1} \epsilon_n v_n.
\end{equation}
This formula assumes that the impact $\lambda \epsilon_n v_n$ of trades in the n$^{th}$ time interval persists unabated up to time $T$. Huberman and Stanzl (2004), and Farmer (2002) show that the linear price adjustement of the Kyle model is the only specification that does not allow price manipulations, {\it provided impact is permanent}, see below. From the above equation, it is clear that the sign of the trades must be serially uncorrelated if the price is to follow an unpredictable random walk. Within the setting of the Kyle model, the trading schedule of the insider is precisely such that the $\epsilon_n$ are uncorrelated (Kyle, 1985). But data from real markets reveal correlations of the sign of the traded volume over long time scales; we will come back to this important point below.

\section{Measures of price impact}

An interesting measure of price impact is the correlation between the price change between $0$ and $T$ and the sign of the trade at time $0$, defined
in general as:
\begin{equation}
R(T) = E\left[(p_T-p_0) \cdot \epsilon_0\right] - E\left[(p_T-p_0)\right]E\left[\epsilon_0\right].
\end{equation}
In the following, we will assume that drifts can be neglected so that only the first term in the RHS is not zero. This is often the
case in practice if $T$ is chosen to be sufficiently small and/or if there is no strong buy/sell assymetry ($E\left[\epsilon_0\right]=0$).

Within the Kyle framework where the $\epsilon$s are uncorrelated, $R(T)$ is easily computed and is found to be:
\begin{equation}
R(T) = \lambda \, E[v]
\end{equation}
The impact function $R(T)$ is therefore {\it lag independent} in this model. One can of course also define a volume dependent impact function by conditioning the above average on the incoming volume, i.e.:
\begin{equation}
R(T,v) = E\left[(p_T-p_0) \cdot \epsilon_0|v_0=v\right],
\end{equation}
which, in the Kyle model, is equal to $\lambda v$, for all $T$. Note that both $R(T)$ and $R(T,v)$ a priori depend on choice of the elementary time scale $\Delta t$, but
we want to avoid heavy notations and skip this extra variable dependence.   

Another commonly used measure of impact is the correlation $\rho(T)$ between the price change from $0$ to $T$ and the total signed volume in the same interval, or more precisely:
\begin{equation}\label{rho}
\rho(T) = \frac{E\left[(p_T-p_0) \cdot \sum_{n=0}^{N-1} \epsilon_n v_n\right]}{\sqrt{E\left[(p_T-p_0)\right]^2 E\left[(\sum_{n=0}^{N-1} \epsilon_n v_n)^2\right]}}
\end{equation}
In the Kyle model this correlation is trivially equal to unity, but this correlation can decrease if liquidity fluctuates ($\lambda$ becomes time dependent), or if other sources of volatility are taken into account. For example, the MM can revise his quotes in the absence of any trades, say because
of some incoming news. This can be modelled by adding an extra term in Eq. (\ref{evolution}) above:
\begin{equation}\label{evolution1}
p_T = p_0 + \lambda \sum_{n=0}^{N-1} \epsilon_n v_n + \sum_{n=0}^{N-1} \eta_n,
\end{equation}
where the $\eta_n$'s are uncorrelated increments, representing causes of price variations not directly related to trading. 

It is often useful to generalize the above definition of $\rho(T)$ by replacing the volumes $v_n$ by $v_n^\psi$, where $\psi$ is a certain exponent. As will be discussed below, the measured correlations are 
found to be stronger when $\psi < 1$.

\section{Empirical facts: impact cannot be so simple}

Here we summarize briefly the salient empirical facts that emerged in the last twenty years, concerning the volume dependence and the temporal behaviour of impact (see Bouchaud et al. 2009 for a recent review). Note that in order to characterize impact empirically, one has to specify two time scales: a) the ``elementary'' time scale $\Delta t$ over which trades are aggregated; b) the time scale $T$ over which the impact of the initial trade is measured.
 
\subsection{Concave volume dependence}

The Kyle model assumes linear dependence of impact on the traded volume. One can measure the volume dependent {\it instantaneous} impact function 
$R(T=\Delta t,v)$ for different elementary time scales $\Delta t$, ranging from the average transaction time to several hours or 
days. One typically finds that the volume dependence of this impact is {\it sublinear} and well described by a power-law:
\begin{equation}
R(T=\Delta t,v) \propto v^{\psi(\Delta t)}; \qquad \psi(\Delta t) \leq 1.
\end{equation}
The exponent $\psi$ increases with the elementary time scale, taking rather small values $\psi \simeq 0.1 - 0.3$ for individual trades, and increasing 
towards $\psi=1$ when $\Delta t$ corresponds to several thousands of trades, with some concavity remaining for large volumes (Hasbrouck \& Seppi (2001), Plerou et al. (2002)). The correlation coefficient $\rho(T=\Delta t)$ similarly increases as $\Delta t$ increase, and reaches rather large values $> 0.5$ for daily returns of individual stocks, futures or currencies (see Evans and Lyons, 2002).

Note that the small value of $\psi$ at the individual trade level means that impact is only weakly dependent on volume, in line with many studies that show a stronger correlation of price changes with the {\it number of trades} than with the traded volume, see e.g. Jones et al. (1996). The small value of $\psi$ is often interpreted in terms of discretionary trading: large market orders are only submitted when there is a large prevailing volume at the best quote, a conditioning that mitigates the impact of these large orders. 

The above results are established using the total aggregated signed volumes, independently of the origin of the trades. Large brokerage or trading firms can also measure the price impact of their own trades; a concave impact function is usually observed with a value of $\psi$ close to $1/2$, see e.g. Almgren et al. (2005), and 
{\bf Execution Costs}. For example, the BARRA price impact model posits that, on a time interval $\Delta t$ needed to complete a typical trade,
\begin{equation}
R(\Delta t,v) = A \sigma \sqrt{\frac{v}{V}},
\end{equation}
where $\sigma$ is the volatility and $V$ the traded volume per unit time, and $A$ a numerical coefficient of order unity (see BARRA, 1997). Different theoretical
justifications for this square-root impact law are given in BARRA (1997), Grinold \& Kahn (1999) and Gabaix et al. (2006).

In the case of stocks, one can also study empirically the influence of the market capitalisation $M$. One finds that when $\Delta t$ corresponds 
to a single trade, the data can be approximately rescaled as (Lillo et al. 2003):
\begin{equation}
R(\Delta t,v)  \approx M^{-0.3} F\left(M^{0.3}\frac{v}{\overline v}\right),
\end{equation}
where $\overline v$ is the average volume per trade for a given stock, and $F(.)$ a master function that behaves as a power-law with exponent $\psi$. 

\subsection{Impact cannot be permanent}

As we observed above, a permanent impact model leads to unpredictable price changes only if the signed volume is uncorrelated. However, empirical 
data shows that on a large variety of markets the autocorrelation of the signs $\epsilon_n$ decays extremely slowly with time, over at least 
several days, representing thousands of trades or more (see e.g. Bouchaud et al. 2009). The order flow is therefore found to be strongly persistent and predictable. This comes from the fact that even ``highly liquid" markets only offer very small volumes for immediate execution. The fact that the outstanding liquidity is so small has an immediate consequence: trades must be fragmented, and need several hours, days or even weeks to be completed. This clearly creates long memory in the sign of the order flow and shows that private information can only be {\it slowly} incorporated into prices. This observation, however, is incompatible with a permanent impact model such as Eq. (\ref{evolution}), which would lead to trends, i.e. strongly autocorrelated price changes. 

\subsection{A non-linear, transient impact model}

In order to reconcile persistent order flow with nearly unpredictable price changes, one can postulate a non-linear, transient impact model that
generalizes Eq. (\ref{evolution}) above:
\begin{equation}\label{evolution3}
p_T = p_{-\infty} + \lambda \sum_{n=-\infty}^{N-1} G(N-n) \epsilon_n v_n^\psi,
\end{equation}
where $G(\ell)$ describes the temporal behaviour of impact. One can show that it is possible to choose a certain {\it decaying shape} for the impact function $G(\ell)$ such as to offset exactly the autocorrelation of the order flow and ensure that the price changes are white noise. 

Assume for simplicity that volumes are all equal: $v_n \equiv v, \forall n$. One can then show that if $C(\ell) = E[\epsilon_n \cdot \epsilon_{n+\ell}]$ decays for large 
$\ell$ as $\ell^{-\gamma}$ with $\gamma < 1$ (typically $\gamma \approx 0.5$ for stocks), then $G(\ell)$ should itself decay to zero as $\ell^{-\beta}$
with $\beta=(1-\gamma)/2$ (Bouchaud et al. 2004). A permanent impact component $G(\ell \to \infty) > 0$ is only compatible with the random nature of prices if $C(\ell)$ decays fast enough ($\gamma > 1$). Within this transient impact model with fixed volume of trades, the relation between the price impact function $R(T)$, $G(\ell)$ and $C(\ell)$ reads (Bouchaud et al. 2004):
\begin{equation} \label{response}
R(T=\ell \Delta t) = \lambda v^\psi\left[G(\ell)+ \sum_{0 < j < \ell} G(\ell-j) C(j) + \sum_{j > 0} \left[G_0(\ell+j)-G_0(j)\right] C(j)\right].
\end{equation}
In other words, the impact  $G(\ell)$ of a single trade in isolation is different from the directly measurable impact $R(T)$, which picks up contributions
from the fact that trades tend to repeat themselves in the same direction.

Note that even if the impact $G(\ell)$ of an individual trade decays with time, one can show that 
both the total impact $R(T)$ and the correlation $\rho(T)$ defined by Eq. (\ref{rho}) tend to a non-zero limit when $T$ is large whenever the relation  $\beta=(1-\gamma)/2$ holds.

Finally, we mentioned above that the linear impact model, corresponding to $\psi = 1$, is the only choice consistent with the absence of price manipulation strategies if 
impact is permanent ($\beta=0$). But if impact is transient, other values of $\psi \leq 1$ become acceptable. Gatheral has recently shown that if $\beta+\psi \geq 1$,
price manipulation is not possible (Gatheral, 2009).

\subsection{Another point of view: surprise in the order flow}

If one insists {\it a priori} that prices must follow a strict random walk, then only the {\it surprise} in the order flow can impact the price.
In other words, the impact of the trades in the $n^{th}$ interval of time $\Delta t$ should read (neglecting volume fluctuations):
\begin{equation}\label{MRR}
\Delta p_n = \lambda v^\psi \big(\epsilon_n - E[\epsilon_{n}|I_{n-1}]\big),
\end{equation}
where $I_{n-1}$ is the information set available just before the $n^{th}$ interval of time. 

If the $\epsilon_n$'s are independent, then $E[\epsilon_{n}|I_{n-1}]=0$ and one recovers the specification of the Kyle model. If on the other hand the $\epsilon_n$'s are correlated, one can form a prediction for the next value of $\epsilon_n$ based on the past realisations $\epsilon_{m < n}$, such that the surprise 
component $\epsilon_n - E[\epsilon_{n}|I_{n-1}]$ is by construction uncorrelated for different times. 

Within this simplified framework, 
there are only two possible outcomes: either the sign of the $n$th transaction matches the sign of the predictor $E[\epsilon_{n}|I_{n-1}]$, 
or the signs are opposite. It is easy to show that the more likely outcome, i.e. $\epsilon_n = {\rm{sign}} (E[\epsilon_{n}|I_{n-1}])$, has the smaller impact (Gerig 2007). Because of the positive correlation in order flow, the impact of a buy following a buy should be less than the impact of a sell following a buy -- otherwise trends would appear. The detailed microstructural mechanism for such an history dependent asymmetric impact is a topic of research, see Bouchaud et al. (2009) and references therein.

Using a linear autoregression model for $E[\epsilon_{n}|I_{n-1}]$ allows one to identify the above surprise model, Eq. (\ref{MRR}), with the transient impact model
of the previous section. Following Hasbrouck's VAR model (Hasbrouck, 1991), one may assume that:
\begin{equation}
\label{arModel}
E[\epsilon_{n}|I_{n-1}] =\sum_{j=1}^\infty a_j \epsilon_{n - j},
\end{equation}
where the coefficents $a_j$ can be computed from the sign autocorrelation $C(\ell)$ using standard methods in filtering theory. Then the above transient impact 
model is precisely recovered provided the following identification holds:
\begin{equation}
G(\ell)=1- \sum_{j=1}^{\ell-1} a_j.
\end{equation}

\section{Spread and impact are two sides of the same coin}

Since market makers (or liquidity providers) cannot guess the surprise of the next trade, they post a bid price $b_n$ and an ask price $a_n$ given by:
\begin{equation}
a_n = p_{n-1} + \lambda v^\psi \left(1 - E[\epsilon_{n}|I_{n-1}]\right); \qquad b_n = p_{n-1} + \lambda v^\psi \left(-1 - E[\epsilon_{n}|I_{n-1}]\right),
\end{equation}  
The above rule ensures no {\it ex-post} regrets for the market maker: whatever the sign of the trade, the traded price is
always the `right' one (see Madhavan et al. (1997)). The bid-ask spread $S$ is then given by:
\begin{equation}
S = a_n - b_n = 2 \lambda v^\psi,
\end{equation}
showing that spread and impact are two sides of the same coin: liquidity providers must be compensated for the adverse impact of market order 
trades (see {\bf Adverse Selection, Limit Order Markets}). Conversely, one sees that the impact of trades is expected to be proportional to the bid-ask spread, suggesting that the volatility per trade $\sigma_1$ is also proportional to the bid-ask spread. Such a correlation is well supported by empirical data, 
see Wyart et al. (2008). The relation with the volatility {\it per unit time} $\sigma$ involves the trading frequency $f$, through $
\sigma =  \sigma_1 \sqrt{f}$.

\section{Conclusion}

Although ``price impact'' seems to convey the idea of a forceful and intuitive mechanism, the story behind it might not be that simple. Empirical studies show that
the correlation between signed order flow and price changes is indeed strong, but the impact of trades is neither linear in volume nor permanent, as assumed
in several models, such as the Kyle model. Impact is rather found to be strongly concave in volume and transient, the latter property being a necessary consequence
of the long-memory nature of the order flow. Only after averaging on a long time scale (on the order of days) may an {\it effective} linear and permanent model
make sense. This is an important observation for execution costs monitoring and control, but also for building agent-based models of market activity that often 
posit linear and permanent impact. 

Coming back to Hasbrouck's comment (Hasbrouck 2007), do trades {\it impact} prices or do they {\it forecast} future price changes? Since trading on modern electronic markets is anonymous,  there cannot be any obvious difference between ``informed'' trades and ``uninformed'' trades if the strategies used for their execution are similar. Hence, the impact of any trade must statistically be the
same, whether informed or not informed. Impact indeed allows private information to be reflected in prices, but by the same token, random fluctuations in order flow (induced by noise traders) must also contribute to the volatility of markets.

\section*{References}

$ $

R. Almgren, C. Thum, H. L. Hauptmann, and H. Li. {\it Equity market impact}, Risk, July 2005.

BARRA, {\it Market impact model handbook} (Berkeley, California, Barra, 1997).

J.-P. Bouchaud, Y. Gefen, M. Potters, and M. Wyart. {\it Fluctuations and response
in financial markets: The subtle nature of "random" price changes}, Quantitative Finance, 4(2):176-190 (2004)

J.-P. Bouchaud, J. D. Farmer, F. Lillo, {\it How markets slowly digest changes in supply and demand}, in: Handbook of 
Financial Markets: Dynamics and Evolution, North-Holland, Elsevier, 2009.

M. Evans, R. Lyons, {\it Order Flow and Exchange Rate Dynamics}, Journal of Political Economy, 110, 170-180 (2002)

J. D. Farmer, {\it Market force, ecology and evolution}, Industrial and Corporate Changes, 11, 895-953 (2002)

J. D. Farmer, P. Patelli, I. Zovko, {\it The predictive power of zero intelligence in financial markets}, PNAS, 102, 2254-2259 (2005)

X. Gabaix, P. Gopikrishnan, V. Plerou, and H. Stanley, {\it Institutional investors and stock market volatility}, Quarterly Journal of Economics, 121:461-504
(2006).

J. Gatheral, {\it No dynamic arbitrage and market impact}, SSRN, abstract-id=1292353 (2008). 

A. Gerig, {\it A Theory for Market Impact: How Order Flow Affects Stock Price},
PhD thesis, University of Illinois (2007), available at: arXiv:0804.3818

R. C. Grinold, R. N. Kahn, {\it Active Portfolio Management}, McGraw-Hill, 1999.

J. Hasbrouck, {\it Measuring the Information-Content of Stock Trades} Journal of Finance, 46, 179-207 (1991)

J. Hasbrouck, D. Seppi, {\it Common Factors in Prices, Order Flows and Liquidity}, Journal of Financial Economics, 59, 388-411 (2001)

J. Hasbrouck, {\it Empirical Market Microstructure}, Oxford University Press, 2007.

G. Huberman, W. Stanzl, {\it Price Manipulation and Quasi-Arbitrage}, Econometrica, 74, 1247-1276 (2004)

C. Jones, G. Kaul, M. L. Lipson, {\it Transactions, volume, and volatility}, Review of Financial Studies 7,
631-651 (1994)

A. Kyle, {\it Continuous Auctions and Insider Trading}, Econometrica, 53, 1315-1335 (1985)

F. Lillo, J. D. Farmer, and R. N. Mantegna, {\it Master curve for price impact function}, Nature, 421:129-130 (2003)

A. Madhavan, M. Richardson, M. Roomans, {\it Why do Security Prices Fluctuate? A Transaction-Level Analysis of NYSE Stocks}, Review of Financial Studies,
10, 1035-1064 (1997)

V. Plerou, P. Gopikrishnan, X. Gabaix and H. E. Stanley, {\it Quantifying Stock Price Response to Demand Fluctuations}, Physical Review E, 66, 027104 (2002)

M. Wyart, J.-P. Bouchaud, J. Kockelkoren, M. Potters, M. Vettorazzo, {\it Relation between Bid-Ask Spread, Impact and Volatility in Double Auction Markets}, Quantitative Finance, 8, 41-57 (2008). 

\end{document}